\documentclass[11pt,twoside]{article}
\usepackage{./asp2014}

\aspSuppressVolSlug
\resetcounters

\def\actaa{Acta Astron.}      
\def\procspie{Proceedings of the SPIE} 

\begin{document}

\title{Cepheid distances based on Gaia and VMC@VISTA observations}
\author{V. Ripepi$^1$,R. Molinaro$^1$,  M. Marconi$^1$ and  M.-R. L. Cioni$^2$ 
\affil{$^1$INAF-Osservatorio Astronomico di Capodimonte, Naples, Italy; \email{vincenzo.ripepi@inaf.it,roberto.molinaro@inaf.it,marcella.marconi@inaf.it}}
\affil{$^2$ Leibniz-Institut f\"ur Astrophysik Potsdam, An der Sternwarte 16, D-14482 Potsdam, Germany; \email{mcioni@aip.de}
}}


\paperauthor{V. Ripepi}{vincenzo.ripepi@inaf.it}{0000-0003-1801-426X}{INAF-Osservatorio Astronomico di Capodimonte}{}{Naples}{}{I-80131}{Italy}
\paperauthor{M.Molinaro}{roberto.molinaro@inaf.it}{0000-0003-3055-6002}{INAF-Osservatorio Astronomico di Capodimonte}{}{Naples}{}{I-80131}{Italy}
\paperauthor{M.Marconi}{marcella.marconi@inaf.it}{0000-0002-1330-2927}{INAF-Osservatorio Astronomico di Capodimonte}{}{Naples}{}{I-80131}{Italy}
\paperauthor{M.-R.Cioni}{mcioni@aip.de}{0000-0002-6797-696X}{ Leibniz-Institut f\"ur Astrophysik Potsdam}{}{Potsdam}{}{D-14482}{Germany}

\begin{abstract}
We present new accurate Period-Luminosity (PL) and Period-Wesenheit (PW) relations in the V,J,Ks bands based on a sample of more than 4500 Cepheids in the Large Magellanic Cloud (LMC) whose photometry was obtained in the context of the VISTA Magellanic Clouds (VMC) Survey. The excellent precision of these data allows us to study the geometry of the LMC and to establish a solid baseline for extra-galactic distance scale studies.  To calibrate the zero points of these PL/PW relations, we adopted Gaia Data Release 2 parallaxes for more than 2000 Milky Way Cepheids. The implications for the measurement of $H_0$ are briefly discussed.
\end{abstract}

\section{Introduction}

Classical Cepheids (DCEPs) are the most important standard candles for the extragalactic distance scale \citep{Riess2016} thanks to their peculiar period-luminosity (PL) and Period-Wesenheit (PW) relations \citep[e.g.][]{Leavitt1912,Madore1982,Caputo2000}.
Once calibrated with geometrical methods (e.g. trigonometric parallaxes, eclipsing binaries, water maser), these relations can be used to calibrate secondary distance indicators such as Type Ia Supernovae (SNe), whose luminosity are sufficiently bright to allow us to estimate the distances of galaxies in the Hubble flow. A measure of the slope of the relation between the distances of these galaxies and their recession velocity allows us to directly estimate the Hubble constant ($H_0$), one of the most important quantities in astrophysics, as it expresses the rate at which the universe is expanding and its inverse represents the age of the universe. This is actually the so-called distance ladder that has been used for decades to estimate $H_0$ \citep[e.g.][]{Sandage2006,Freedman2012,Riess2016,Riess2018b,Riess2019}.

Recent determinations of $H_0$ using the distance ladder \citep{Riess2016} are in tension with the Planck Cosmic Microwave Background (CMB) measurements under the flat $\Lambda$ Cold Dark Matter ($\Lambda$CDM) model \citep[e.g.][]{Freedman2017}. The latest estimate from the Riess group is $H_0$=74.03$\pm$1.42 km s$^{-1}$ Mpc$^{-1}$ \citep[][]{Riess2019} is in tension with the Planck flat $\Lambda$CMB result of $H_0$=67.4$\pm$0.5 km s$^{-1}$ Mpc$^{-1}$ \citep{Planck2018} by 4.4$\sigma$. 

This tension has been confirmed using different methods to estimate the local value of $H_0$: i) adopting other primary distance indicators instead of DCEPs: Miras \citep{Huang2019} and Tip of the Red Giant Branch \citep[TRGB,][]{Yuan2019}; ii) using DCEPs but with improved distances of the Large Magellanic Cloud (LMC) and NGC4258, the two fundamental anchors of the distance ladder \citep{Reid2019}; iii) using the gravitational time-delays of six lensed quasars \citep[H0LICOW and STRIDES collaborations][]{Wong2019}\footnote{https://shsuyu.github.io/H0LiCOW/site/; http://strides.astro.ucla.edu/}. 
The latter method is completely independent from the distance ladder, suggesting that the tension is real. Therefore the challenge now is to reduce the errors (random and systematic) of each method in order to characterize the tension with an accuracy sufficient to investigate what cosmological models can explain it.

Concerning the distance ladder, one of the main sources of uncertainties are the actual values of the slopes and intercepts of DECP PL/PW relations and their dependence on parameters such as the dispersion of the instability strip, metallicity, duplicity etc. 
Since the intrinsic dispersion of  the PL relation, due to the strip topology, strongly decreases as  filter  wavelength increases, Near-Infrared (NIR) PL relations are commonly used for the distance scale. NIR bands are also much less affected by reddening than in the optical. On the other hand, both in the optical and NIR, the use of PW relations has many advantages, as they are reddening free by construction, and the color term included in their definition takes partially into account the width of the instability strip, thus reducing significantly their intrinsic dispersion compared to PL relations. 
However, both PL and PW relations show a dependence on metallicity
that, although smaller in the NIR, has to be taken into account to avoid systematic effects in the determination of distances \citep[][and references therein]{Romaniello2008,Bono2010,Gieren2018}. 

The PW relations actually used in the distance ladder by  Riess' group are calibrated using a small number of DCEPs with accurate Hubble Space Telescope (HST) parallaxes \citep{Riess2018a}, and the procedure takes into account the metallicity. 
However, there is still not a general consensus about the actual extent of the magnitude dependence on metallicity of the slope and intercept of the PL/PW relations in the different bands \citep[e.g.][]{Macri2006, Romaniello2008,Bono2010,Freedman2011,Shappee2011,Pejcha2012,Kodric2013,Fausnaugh2015,Riess2016}. This is mainly because, with the exception of the Magellanic Clouds DCEPs \citep{Romaniello2008}, most measurements relies on DCEPs hosted in distant galaxies, whose metallicities are known with low precision. A direct measurement using Galactic DCEPs with well known [Fe/H] estimates based on high-resolution spectroscopy was hampered until recent times by the lack of accurate independent distances for the Galactic DCEPs \citep{Groenewegen2018,Ripepi2019}. 

\articlefigure[width=.7\textwidth]{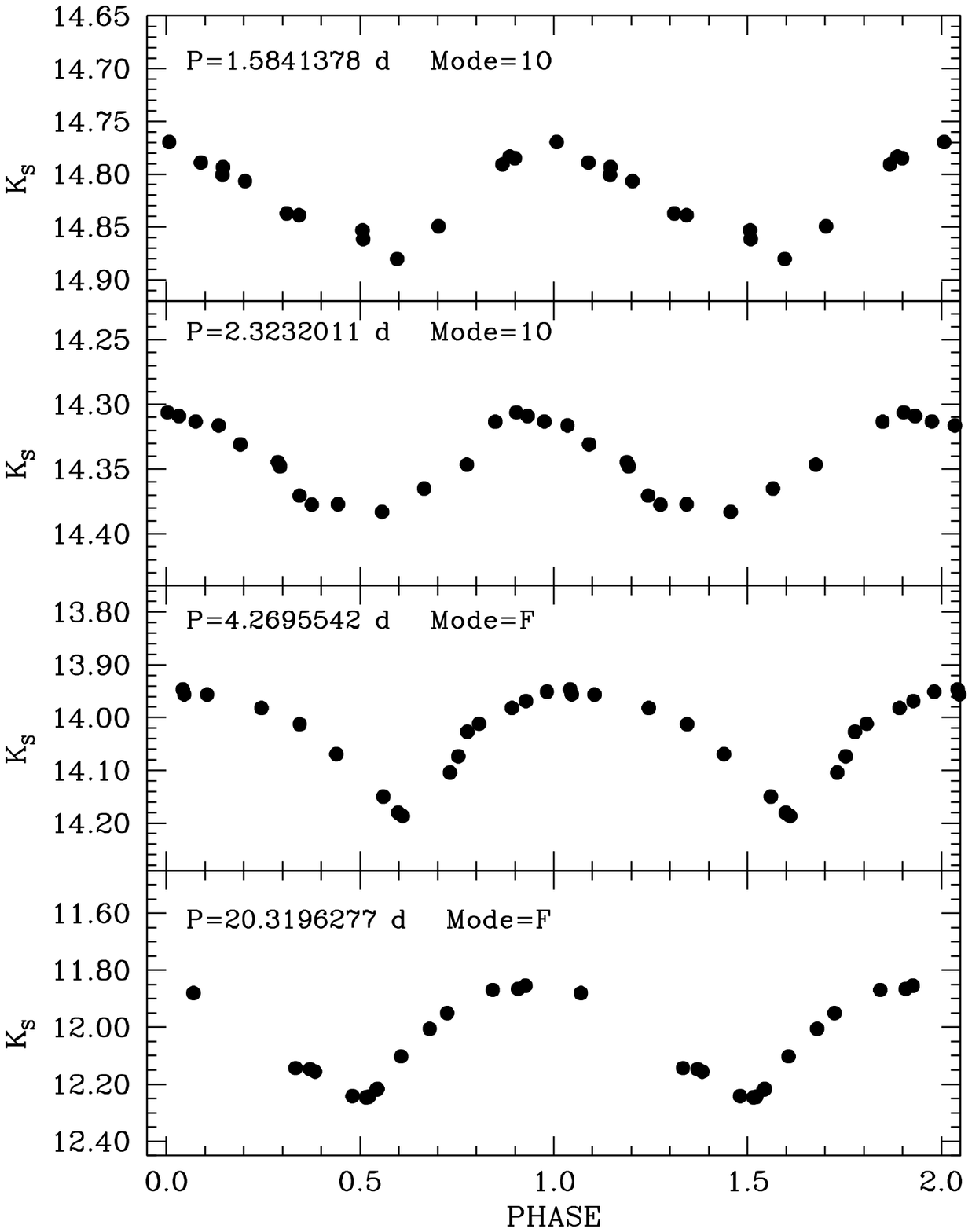}{fig:lc}{Typical VMC light curves for F and 1O DCEPs.}

In this context, the purpose of this project is to try to reduce the uncertainties inherent to the use of DCEPs as standard candles. In particular, we intend to take advantage of the precise NIR $J,K_s$ photometry from the Vista Magellanic Cloud survey \citep[VMC,][]{Cioni2011} to obtain better PL/PW for DCEPs in the LMC. Moreover, we aim at using the Gaia satellite Data Release 2 \citep[DR2,][]{GaiaPrusti2016,GaiaBrown2018} to check the local calibration of slopes and intercepts of the PL/PW relations used for DCEPs. These data will allow us to check whether the distance of the LMC obtained from DCEPs PL/PW relations is in agreement with the accurate geometric estimate by \citet{Pietrzynski2019}.
To investigate the dependence on metallicity of DCEP PL/PW relations, we started a program to obtain high-resolution spectroscopy for 100 DCEPs using different telescopes/instruments. This project is at an early stage and will not be discussed in this work.  

The paper is organized as follows: in Sect. 2 we present the results from the VMC survey; in Sect. 3 we describe the Galactic DCEPs sample; in Sec. 4 we discuss the calibration of the Galactic DCEPs PW relations; in Sect. 5 we discuss the results.

\section{VMC observations}

The VMC project is a European Southern Observatory (ESO) public survey   \citep[][]{Cioni2011} carried out in the NIR with the 4.1m Visible and Infrared Survey Telescope for Astronomy \citep[VISTA,][]{Sutherland2015}, equipped with VIRCAM \citep[VISTA InfraRed Camera,][]{Dalton2006}. The properties of the pulsating stars observed by the VMC survey in the Magellanic System have been discussed in a series of papers \citep{Ripepi2012a,Ripepi2012b,Moretti2014, Ripepi2014,Ripepi2015,Ripepi2016,Ripepi2017} and are not repeated here.
The data analysed in this work were reduced and calibrated with the VISTA
Data Flow System (VDFS) pipeline v1.3 \citep{Irwin2004,Gonzalez2018}, whereas the time--series were downloaded from the VISTA Science Archive \citep[VSA,][]{Cross2012}. 

The sample of DCEPs in the LMC studied here were taken from the OGLE IV  
collection \citep[][and references therein]{Sos2019} which provide the identification of the objects as well as the periods and V,I photometry. Cross-matching the VMC and OGLE IV data with 0.5\arcsec tolerance, we ended up with a sample of 4560 DCEPs with data in the VMC. More precisely, the sample was composed by 2413 fundamental (F), 1715 first-overtone (1O), 93 F/1O and 303 1O/2O DCEPs, respectively. Typical examples of light-curves are shown in Fig.~\ref{fig:lc}.

We estimated the $J$ and $K_s$ intensity-averaged magnitudes and the peak-to-peak amplitudes for the full sample of 4560 DCEPs using a technique similar to that devised in \citet{Ripepi2016}. In brief, we used our best light curves to build several templates in $J$ and $K_s$ and used a modified
$\chi^2$ technique to identify the best-fitting template (full details will be provided in a forthcoming paper, Ripepi et al. in preparation).

The intensity--averaged magnitudes thus obtained were used to construct 
new $PL$ and $PW$ relations shown in Fig.~\ref{fig:pl}. Note that 
lacking accurate individual reddening estimates, we have used a common value $E(V-I)$=0.08 mag for all the stars. Finally, the PL/PW relations have been calculated by means of standard least--squares fit, adopting a 
$\sigma$-clipping algorithm, with $\sigma$=3.5. The resulting relationships are shown in Fig~\ref{fig:pl} as solid lines for both F and 1O pulsators. We report the $PW(J,K_s)$ relations that are functional for the following discussion in analytical form in the first two lines of Table~\ref{table:results}.  However, note that a new Cambridge Astronomy Survey Unit (CASU) data release (v1.5) will be available soon, and the final coefficients of these relations might change. 


\noindent 
 
These relations are, as far as we know, the most accurate PW for DCEPs in the LMC. However, to be useful in the distance scale context, they need to be calibrated in absolute magnitudes. To this goal we will use Galactic DCEPs in conjunction with Gaia parallaxes.

\articlefiguretwo{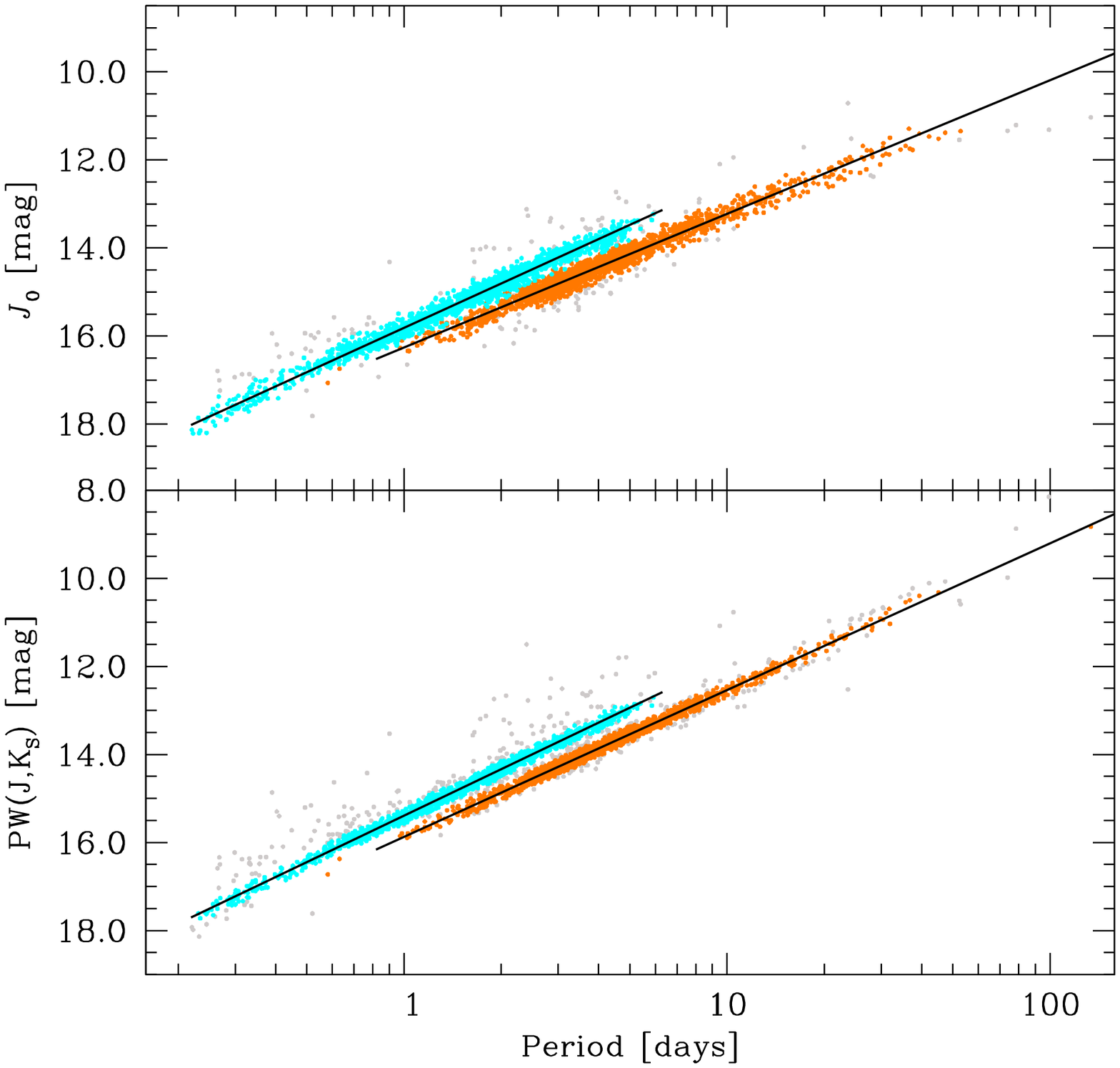}{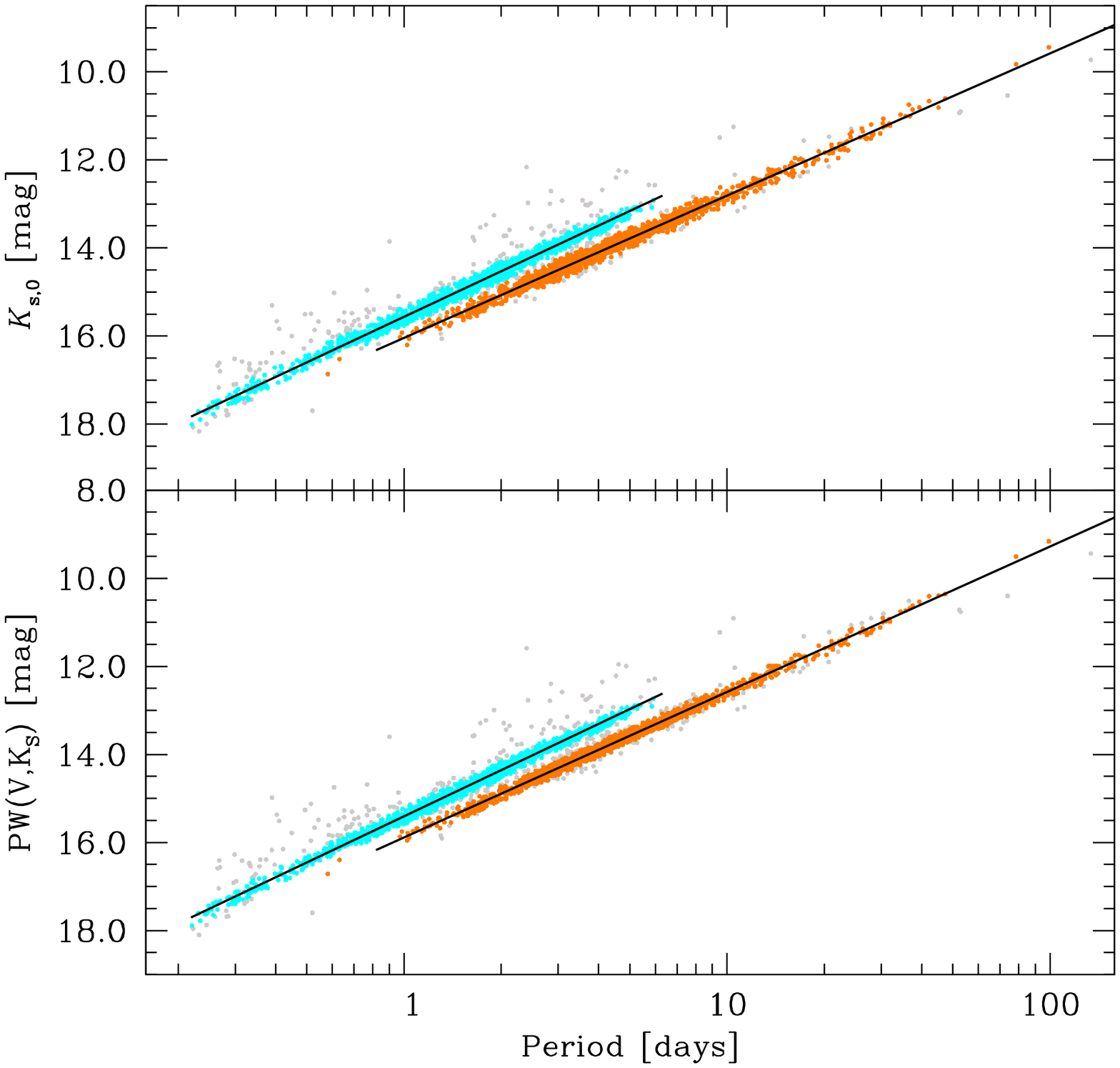}{fig:pl}{\emph{Left:} PL in $J_0$ and $PW(J,K_S)$ for the LMC DCEPs. The Wesenheit magnitude is defined as  $K_s-0.69\times(J-K_s)$. \emph{Right:} The same but for $K_{s,0}$ and $PW(V,K_S)$. In this case, the Wesenheit magnitude is defined as  $K_s-0.13\times(V-K_s)$. In both panels orange and cyan symbols represent F and 1O DCEPs, respectively. The grey points represent the objects not considered in the derivation of the best-fitting regression lines that are shown with solid black lines.}

\section{Galactic DCEPs}

Until a few years ago less than about 800 Galactic DCEPs were known and only about 450 objects were well characterized \citep[e.g.][and references therein]{GaiaClementini2017,Groenewegen2018}.  
The advent of the {\it Gaia} satellite is providing an unprecedented contribution to this field. Indeed, the recent DR2 \citep{GaiaBrown2018} published accurate light curves and astrometry for more than 1000 DCEPs \citep{Clementini2016,Holl2018,Clementini2019}. 

In a recent paper, \citet{Ripepi2019} have reanalysed the sample of {\it
 Gaia} DR2 DCEPs, removing contaminants, and retaining a list of 800 bona-fide DCEPs, among which 123 are new discoveries. Additional new DCEPs have been recently found in the context of other projects, namely: ASAS-SN \citep[All-Sky Automated Survey for Supernovae,][]{Jayasinghe2018}; ATLAS \citep[Asteroid Terrestrial-impact Last Alert System,][]{Heinze2018}; WISE \citep[Wide-field Infrared Survey Explorer,][]{Chen2018}, and OGLE Galactic Disk survey \citep{Udalski2018}, leading the total number of known or candidate Galactic DCEPs to more than 3000 objects.  
A significant fraction of these DCEPs possess astrometric parallaxes from {\it Gaia}. This quantity can be used in conjunction with other means to remove from the sample the contaminants, mainly constituted by binary systems and rotational variables. After the cleaning process (whose details are omitted for brevity), we have a sample of 2164 and 598 bona fide DCEP F and 1O, respectively.

To build PL/PW relations for these objects, periods and multiband photometry are needed. Given the heterogeneity of this sample, these properties are not available for all the objects. The periods are generally available in the paper of discovery, whereas accurate NIR $(J,Ks)$ photometry is available only for about 450 DCEPs \citep[e.g.][]{Groenewegen2018}. For the remaining objects, to estimate the average magnitude in $JK_s$, we used single epoch photometry from 2MASS \citep{Skrutskie2006} and DENIS (http://cds.u-strasbg.fr/denis.html) in conjunction with the template technique \citep{Sos2005}. This method can be used when the ephemerides of the DCEPs are known with reasonable accuracy, and this was true for the vast majority of our sample.  
As for reddening and metallicity, these values are missing for almost any of the newly discovered DCEPs. Therefore, we used Wesenheit magnitudes, that are reddening free by definition, and ignored for the moment the metallicity contribution in building the PW relation.
      
\section{Galactic DCEP PW($J$,$K_S$) relation}

In order to derive the $PW(JK_s)$ we decided to use the Astrometry Based Luminosity \citep[ABL,][]{Arenou1999}, a quantity devised to use the parallax in a linear way, allowing us to include objects with negative parallaxes, thus removing any bias source. The ABL is defined as in Eq.~\ref{eqABL}.  

\begin{equation}
{\rm ABL}=10^{0.2 W(J,K_s)_A}=10^{0.2({\rm b}+{\rm a} \log P)}=\varpi10^{0.2{W(J,K_s)}-2} \qquad
\label{eqABL}
\end{equation}
\noindent
where $W(J,K_S)_A$=a$\log P$+b; $W(J,K_s)_A$ and $W(J,K_s)$ are the absolute and relative Wesenheit magnitudes, respectively. The observables are $W(J,K_s)$, $P,$ and the parallaxes $\varpi$ to which a Zero Point (ZP) offset of 0.046 mas has been applied according to \citet{Riess2018a}.
The unknown $a$ and $b$ values are evaluated by means of a robust weighted least-squares fit procedure whereas their uncertainties are estimated by means of a bootstrap technique. The ABL fit was performed by both varying the slope and fixing it to the value of the LMC. An example of the procedure is shown in Fig.~\ref{fig:abl}, whereas the results of the fitting procedure are reported in Table~\ref{table:results} and discussed in the next section.

\articlefigure[width=1.0\textwidth]{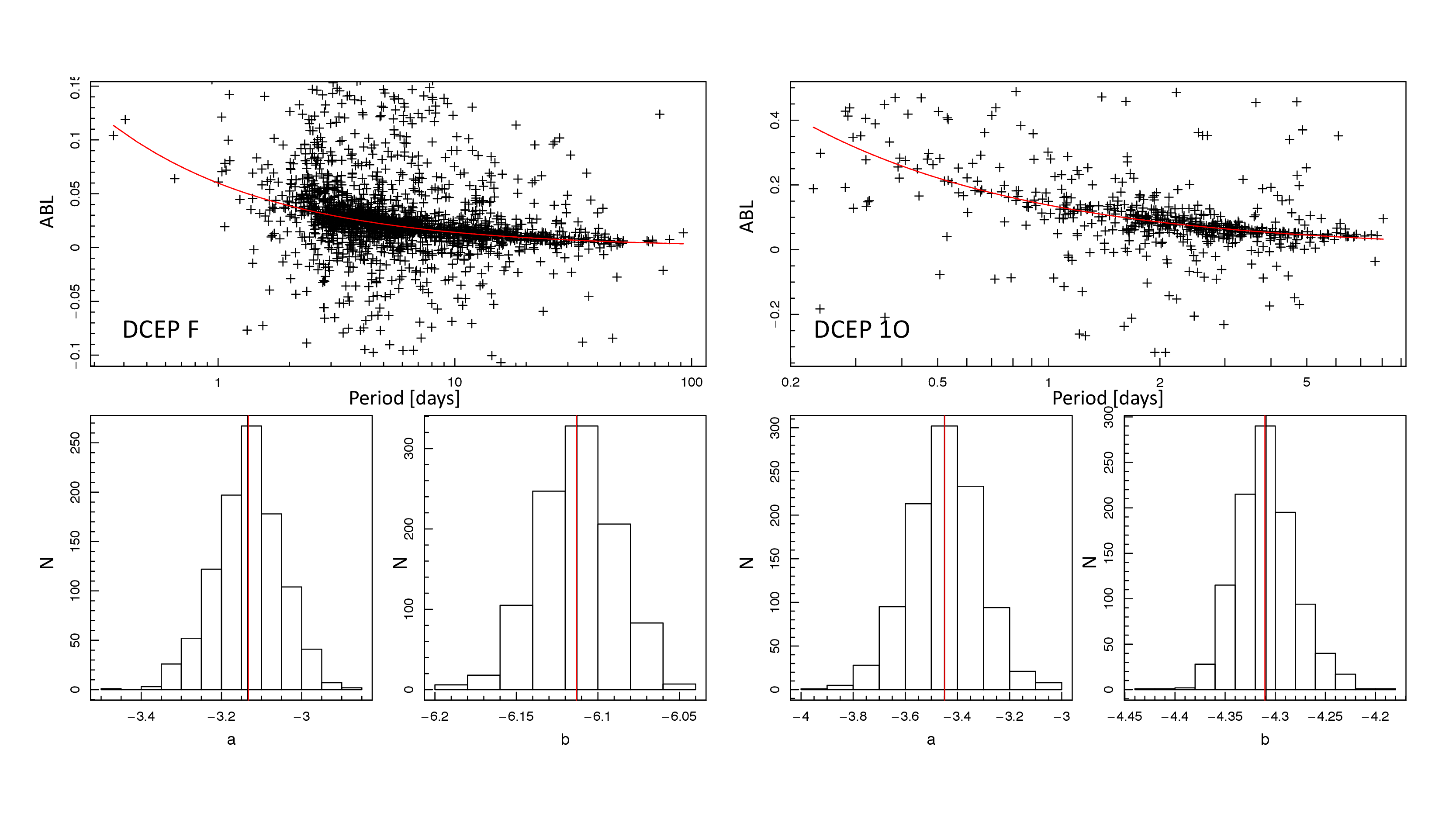}{fig:abl}{Evaluation of the coefficients of the $PW(J,K_s)$ relation through the fit of the ABL function. Top panels show the dependence of the ABL function from period. Black crosses are the DCEPs, red lines are the fit to the data. Bottom panels display the distribution of slope $a$ and intercept $b$ as a result of the bootstrap procedure. The red lines represent the median of the distribution, i.e. the adopted values. Left and right panels show the results for F and 1O DCEPs, respectively.}

\begin{table}[!ht]
\caption{Results of the ABL fitting procedure. The $PW(J,K_s)$ are defined as: $W(J,K_s)=a(logP-1.0)+b$ and $W(J,K_s)=a(logP-0.3)+b$ for F and 1O DCEP, respectively. Note the use of pivoting periods to reduce the correlation between the slope and the intercept of the PL/PW relations.}
\label{table:results}
\smallskip
\begin{center}
{\small
\begin{tabular}{ccccc}  
\tableline
\noalign{\smallskip}
Mode & Galaxy & a & b & DM$_{\rm LMC}$ \\
\noalign{\smallskip}
\tableline
\noalign{\smallskip}
F	& LMC  & -3.332$\pm$0.007 &    12.538$\pm$0.003     & \\
1O	& LMC  & -3.501$\pm$0.007  &   14.340$\pm$0.002     &\\
F   &  MW  &   -3.133$\pm$0.079 &	   -6.113$\pm$0.024 &  \\       
1O  & MW    & -3.443$\pm$ 0.124 &	   -4.310$\pm$0.029&    \\
F	& MW    & -3.332 Fixed	  & -6.155$\pm$0.017   &18.69$\pm$0.02    \\ 
1O  &  MW   &  -3.501 Fixed &	   -4.305$\pm$0.027 &18.64$\pm$0.03   \\
\noalign{\smallskip}
\tableline\
\end{tabular}
}
\end{center}
\end{table}

\section{Discussion}

An inspection of Table~\ref{table:results} reveals that the slopes for LMC and MW of F mode DCEPS are different at a level of 2.5$\sigma$, whereas for 1O DCEPs the slopes agree well within 0.5$\sigma$. This means that some metallicity effect is affecting not only the intercept but also the slope of the $PW(J,K_s) $ relation. However, if we impose the slopes of the LMC to the MW DCEPs, we can directly measure the distance modulus (DM) of the LMC by simply comparing the intercept of the PW relations in lines 5,6 and 1,2 of Table~\ref{table:results}. The resulting DMs for the LMC are shown in the last column of Table~\ref{table:results}. It can be seen that these DMs are longer by about 0.2 mag with respect to the geometric distance of $\sim$18.48 mag by \citet{Pietrzynski2019}. However, we have to take into account the effect of the metallicity that is now all in the intercept, as we have imposed the slope of the LMC.  As remarked previously, the metallicity dependence of the DCEPs PL/PW is uncertain. 
To estimate the impact of the metallicity effect on the $JK$ Wesenheit relation, we adopt a metallicity term $\sim$-0.20 mag/dex following some recent results, \citep[i.e.][]{Gieren2018,Groenewegen2018,Ripepi2019} although its accuracy is limited (slightly less than 1 $\sigma$). Assuming a $\Delta [Fe/H]\sim0.4$ dex
between MW and LMC, we obtain a correction of the order of $\sim$-0.1 mag on the value of DM$_{LMC}$. This means that the DM$_{LMC}$ derived using the {\it Gaia} parallaxes remains larger by $\sim$0.1 mag with respect to 
the reference value, implying that an additional ZP offset of $\sim$0.02 mas should be applied to the {\it Gaia} parallaxes. A similar conclusion was reached by \citet{Groenewegen2018} and \citet{Ripepi2019} using different samples of Galactic DCEPs.
We note that such an uncertainty on the ZP offset has a dramatic impact on the measure of $H_0$ with the distance ladder, since it would imply a proportional uncertainty of about 5\% in $H_0$. 
However, measuring the ZP offset of {\it Gaia} parallaxes is not an easy task, as apparently it depends on the class of objects used to estimate this value as well as their distribution on the sky \citep[see][]{Arenou2018}.
Future data releases of the {\it Gaia} satellite will hopefully fix this problem since it is extremely important to know with an accuracy of 3-4 $\mu$as the parallax ZP offset to reduce the impact of the PL/PW uncertainties on $H_0$ to a negligible level.

\acknowledgements 
This work has made use of data from the European Space Agency (ESA) mission
{\it Gaia} (\url{https://www.cosmos.esa.int/gaia}), processed by the {\it Gaia} Data Processing and Analysis Consortium (DPAC,
\url{https://www.cosmos.esa.int/web/gaia/dpac/consortium}). Funding for the DPAC has been provided by national institutions, in particular the institutions participating in the {\it Gaia} Multilateral Agreement.
In particular, the Italian participation
in DPAC has been supported by Istituto Nazionale di Astrofisica
(INAF) and the Agenzia Spaziale Italiana (ASI) through grants I/037/08/0,
I/058/10/0, 2014-025-R.0, and 2014-025-R.1.2015 to INAF (PI M.G. Lattanzi).
V.R. and M.M. acknowledge partial support from the project "MITiC: MIning The Cosmos Big Data and Innovative Italian Technology for Frontier Astrophysics and Cosmology”  (PI B. Garilli).
MRC acknowledges support from the European Research Council (ERC) under European Union’s Horizon 2020 research and innovation programme (grant agreement No 682115).
We thank M. Groenewegen for a careful reading of the manuscript. 


\end{document}